\newcommand\araa{{ARA\&A\,}}%
\newcommand\apj{{ApJ\,}}%
\newcommand\apjl{{ApJ\,}}%
\newcommand\apjs{{ApJS\,}}%
\newcommand\apss{{Ap\&SS\,}}%
\newcommand\aap{{A\&A\,}}%
\newcommand\aaps{{A\&AS\,}}%
\newcommand\mnras{{MNRAS\,}}%
\newcommand\pasp{{PASP\,}}%
\newcommand{\vhbb}{$\Delta V_{\rm HB}^{\rm Bump}\,$}
\newcommand{\vtob}{$\Delta V_{\rm TO}^{\rm Bump}\,$}

\documentclass[graybox, natbib, footinfo]{svmult}

\usepackage{mathptmx}       
\usepackage{helvet}         
\usepackage{courier}        
\usepackage{type1cm}        
\usepackage{makeidx}         
\usepackage{graphicx}        
\usepackage{multicol}        
\usepackage[bottom]{footmisc}

\usepackage{journals}         
\usepackage{url}
\makeindex             

\begin{document}
\title*{Uncertainties and systematics in stellar evolution models of Red Giant Stars}   
\author{Santi Cassisi}   
\institute{S. Cassisi \at INAF - Astronomical Observatory of Collurania, 
Via M. Maggini, 64100 Teramo, Italy, \email{cassisi@te.astro.it}}  
\maketitle
\abstract{In this last decade, our knowledge of evolutionary and structural properties of stars of different
mass and chemical composition has significantly improved. 
This notwithstanding, updated stellar models are still affected by significant and, usually, not negligible
uncertainties. These uncertainties are related to our poor knowledge of some physical processes occurring in the real stars
such as the efficiency of mixing processes.
These drawbacks of stellar models have to be properly taken into account when comparing theory with observations.
In this paper we briefly review current uncertainties affecting low-mass stellar models, i.e. those structures with mass in the range between
$0.6M_\odot$ and $\sim1.4M_\odot$, during the Red Giant Branch stage.
}

\section{Introduction} 

During the second half of last century, stellar evolution theory has allowed us to 
understand the Color Magnitude Diagram (CMD)
of both galactic globular clusters (GGCs) and open clusters, so that now 
we can explain the distribution of stars in the observed CMDs in terms of the nuclear evolution of
stellar structures and, thus, in terms of cluster age and chemical composition.
In recent years, however, the impressive improvements achieved for both photometric
and spectroscopic observations as well as asteroseismological measurements, has allowed us to collect
data of an un-precedent accuracy, which provide at the same time a stringent test and a challenge 
for the accuracy of the models.

On the theoretical side, significant improvements have been achieved
in the determination of the Equation of State (EOS) of the stellar
matter, opacities, nuclear cross sections, neutrino emission rates,
that are, the physical inputs needed in order to solve the equations of
stellar structure. 

The capability of current stellar models to account for
all the observed evolutionary phases is undoubtedly 
an exciting achievement  which crowns with success the development of 
stellar evolutionary theories as pursued all along the second half of the last
century. Following such a success, one is often tempted to use
evolutionary results in an uncritical way, i.e., taking these results at
their face values without accounting for theoretical uncertainties. 
However, these uncertainties do exist, as it is clearly shown by the
not negligible differences still existing among evolutionary results 
provided by different research groups \citep[see the discussion in][]{Chaboyer95, Cassisi98, Cassisi99, Castellani99, Cassisi04}. 

We will discuss the main \rq{ingredients}\rq\ necessary for
computing stellar models and show how the uncertainties on these inputs affect
theoretical predictions of the evolutionary properties of Red Giant Branch (RGB) low-mass stars.

\section{Stellar evolution: the ingredients} 

The stellar structure equations are well known since long time, and a clear description of the physical meaning of each one 
of them can be found in several books \citep[as, for instance,][]{Kippenhahn90}.

The (accurate) numerical solution of these differential equations is no longer a problem and 
it can be easily and quickly achieved when using modern numerical solution schemes and current
generation of powerful computers. This notwithstanding, in order to solve these equations, boundary conditions have to be provided: the boundary condition at the stellar centre are trivial \citep[see]{Salaris05}; however the same does not apply for those at the stellar surface, i.e.,  the
values of temperature and pressure at the base of the atmosphere. These boundary conditions can be
obtained either by adopting an empirical relation for the thermal stratification like that provided
by \cite{Krishna66} or a theoretical approximation as the so-called Eddington approximation.
A more rigorous procedure is to use results from model atmosphere computations  \citep{VandenBerg08}. 

In order to compute a stellar structure, it is fundamental to have an accurate description of the
physical behaviour of the matter in the thermal conditions characteristics of the stellar interiors
and atmospheres. This means that we need to know several physical
inputs as: opacity, EOS, nuclear cross-sections, neutrino energy losses.
A rich literature exists describing the improvements which have been achieved in this
last decade concerning our knowledge of these physical inputs \citep[][and references therein]{Catelan09, 
Cassisi98,Cassisi99, Salaris02}.

Some important assumptions have also to be made concerning the
efficiency of those mechanisms, such as  atomic diffusion and radiative levitation, which can modify the chemical stratification in the interiors and atmosphere.
Until few years ago, all these non-canonical processes were usually ignored in stellar models
computations. However, helioseismology has clearly shown how important is to include atomic
diffusion in the computation of the so-called Standard Solar Model (SSM), in order to obtain a good
agreement between the observed and the predicted frequencies of the non-radial p-modes \citep{jcd93}.
 In the meantime, quite recent spectroscopical
measurements for low-mass, metal-poor stars in GGCs 
strongly point out the importance of including radiative levitation in stellar computations in
order to put in better agreement empirical estimates with the predictions provided by diffusive
models.

When dealing with stellar model computations, one has also to account for the occurrence of mixing. Due to the poor knowledge of how to manage the mixing processes in a stellar
evolutionary code, the efficiency of convection is commonly treated by adopting some approximate
theory. In this context, it has to be noticed that when treating a region where convection is stable,
one has to face with two problems: i) What is the \lq{right}\rq\ temperature gradient in such region?,
ii) What is the \lq{real}\rq\ extension of the convective region?

The first question is really important only when considering the outer convective regions such as
the convective envelopes of cool stars. This occurrence is due to the evidence that, in the stellar
interiors as a consequence of the high densities and, in turn, of the high efficiency of energy transport by convective motions, the \lq{real}\rq\ temperature gradient has to be equal to the
adiabatic one. This consideration does not apply when considering the outer, low-density, stellar
regions, where the correct temperature gradient has to be larger than the adiabatic one: the
so-called {\sl superadiabatic gradient}. One of the main problem in computing star
models is related to the correct estimate of this superadiabatic gradient. 

Almost all evolutionary computations available in literature rely on the mixing length theory \citep[MLT][]{BohmVitense58}. It contains a number of free parameters, whose   
numerical values affect the model $T_{\rm eff}$; one of them  
is $\alpha_{\rm MLT}$, the ratio of the mixing length to the pressure scale   
height, which provides the scale length of the convective motions.   
There exist different versions of the MLT, each one assuming different   
values for these parameters. However, as demonstrated by \cite{Pedersen90} and \cite{Salaris08}, the $T_{\rm eff}$ values obtained from the   
different formalisms can be made consistent, provided that a suitable value of   
$\alpha_{\rm MLT}$ is selected. Therefore, at least for the evaluation of    
$T_{\rm eff}$, the MLT is basically a one-parameter theory.   
 The value of $\alpha_{\rm MLT}$  is usually calibrated by reproducing the solar $T_{\rm eff}$, 
and this solar-calibrated value is then used for computing models 
of stars very  different from the Sun (e.g. metal poor giants). 
It is worth recalling that there exists also an alternative formalism for the computation 
of the superadiabatic   
gradient: the so-called Full-Spectrum-Turbulence theory \citep[FST][]{Canuto96}, a MLT-like formalism with a more sophisticated   
expression for the convective flux, and the scale-length   
of the convective motion fixed a priori. 

For low-mass stars, the problem of the real extension of a convective region really affects only the convective
envelope. In the canonical framework it is assumed that the border of a convective region is fixed
by the condition - according to the classical Schwarzschild criterion - that the radiative gradient
is equal to the adiabatic one. However, it is clear that this condition fixes the point where the
acceleration of the convective cells is equal to zero, so it is realistic to predict that the
convective elements can move beyond, entering and, in turn, mixing the region surrounding the
classical convective boundary. This process is commonly referred to as convective overshoot.
 Convective envelope overshoot could be
important for RGB low-mass stars, since these structures have large convective envelope.

\section{The Red Giant Branch}

The possibility to apply RGB stellar models to fundamental astrophysical problems crucially 
relies on our capability to predict correctly: {\sl i)} the CMD location (in $T_{\rm eff}$ and color) and extension    
(in brightness) of the RGB as a function of the initial chemical composition and age; {\sl ii)}
 the evolutionary timescales all along the RGB; {\sl iii)} the physical and chemical structure of RGB stars.

\subsection{The location and the slope of the RGB}

The main physical inputs used in model computation which affect the RGB location and slope are: the EOS, the low-temperature opacity, the efficiency of superadiabatic convection, and the choice about the outer boundary conditions.

{\sl EOS:} the most recent stellar models rely on updated EOS tabulations, such as the OPAL EOS \citep{Rogers02} and the FreeEOS \citep{Irwin05, Cassisi03}, which allow to properly \lq{cover}\rq\ the whole evolutionary stages. Low-mass RGB models computed by adopting these EOSs are in very good agreement, but the difference increase when comparing also models based on less updated EOS. In this case, one can easily found differences of the the order of $\sim100$~K.

{\sl Radiative opacity:} low-$T$ opacities mainly determine the   
$T_{\rm eff}$ location of theoretical RGB models, while the high-$T$ ones - in particular those for temperature around $10^6K$ - determine the extension of the convective envelope.
 Current generations of stellar models employ mainly the low-$T$ opacity calculations 
by \cite{Alexander94} and by \cite{Ferguson05}, which are the most  up-to-date computations suitable for stellar modeling.     
The main difference between these sets of data and the previous ones is the treatment of molecular absorption, most notably the   
fact that the latest opacity tables include the effect of the several molecules (among
which the $\rm H_{2}O$ that is very important for metal rich RGB stars) and accounts also for the presence of grains. Although significant improvements are still possible as a consequence of a better treatment of the various molecular opacity sources, we do not expect dramatic changes in the temperature regime where the contribution of atoms and molecules dominate. Huge variation can be foreseen in the regime ($T<2000K$) where grains dominates the interaction between radiation and matter. 

When comparing, at different initial metallicities, stellar models\footnote{These stellar models are always based on a solar-calibrated mixing length.}  produced with these two sets of opacities with those based on previous estimates as the \cite{Kurucz93} (K92) ones, one finds that a very good agreement exists   
when $T_{\rm eff}$ is larger than $\sim$4000 K. As soon as the RGB $T_{\rm eff}$ goes below this limit    
(when the models approach the TRGB and/or their initial metallicity is increased),    
the most recent opacity evaluations by \cite{Ferguson05} produce progressively cooler models (differences reaching values of $\approx100$ K or more),  due to the effect of the $\rm H_{2}O$ molecule which contributes substantially to the opacity in this temperature range. 

\begin{figure}
\begin{minipage}[t]{5.75cm}
\begin{center}
\includegraphics[width=5.75cm,clip]{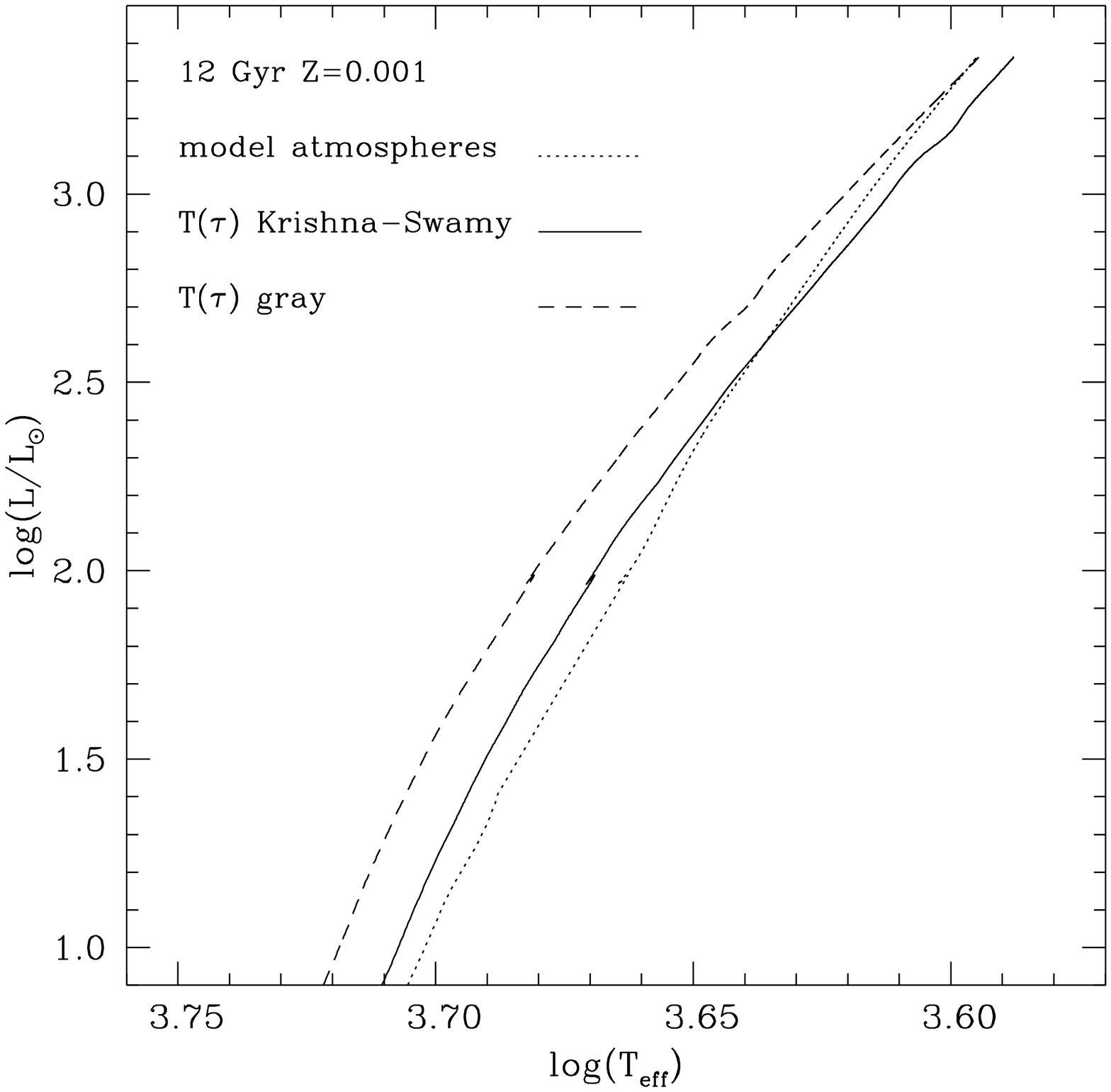}
\caption[Short caption for figure 1]{\label{fig:1} 
The RGB loci of an isochrone computed by adopting different
prescriptions for the outer boundary conditions. All other parameters, as for
instance the $\alpha_{MLT}$ parameter, have been kept fixed.}
\end{center}
\end{minipage}
\hfill
\begin{minipage}[t]{5.75cm}
\begin{center}
\includegraphics[width=5.75cm,clip]{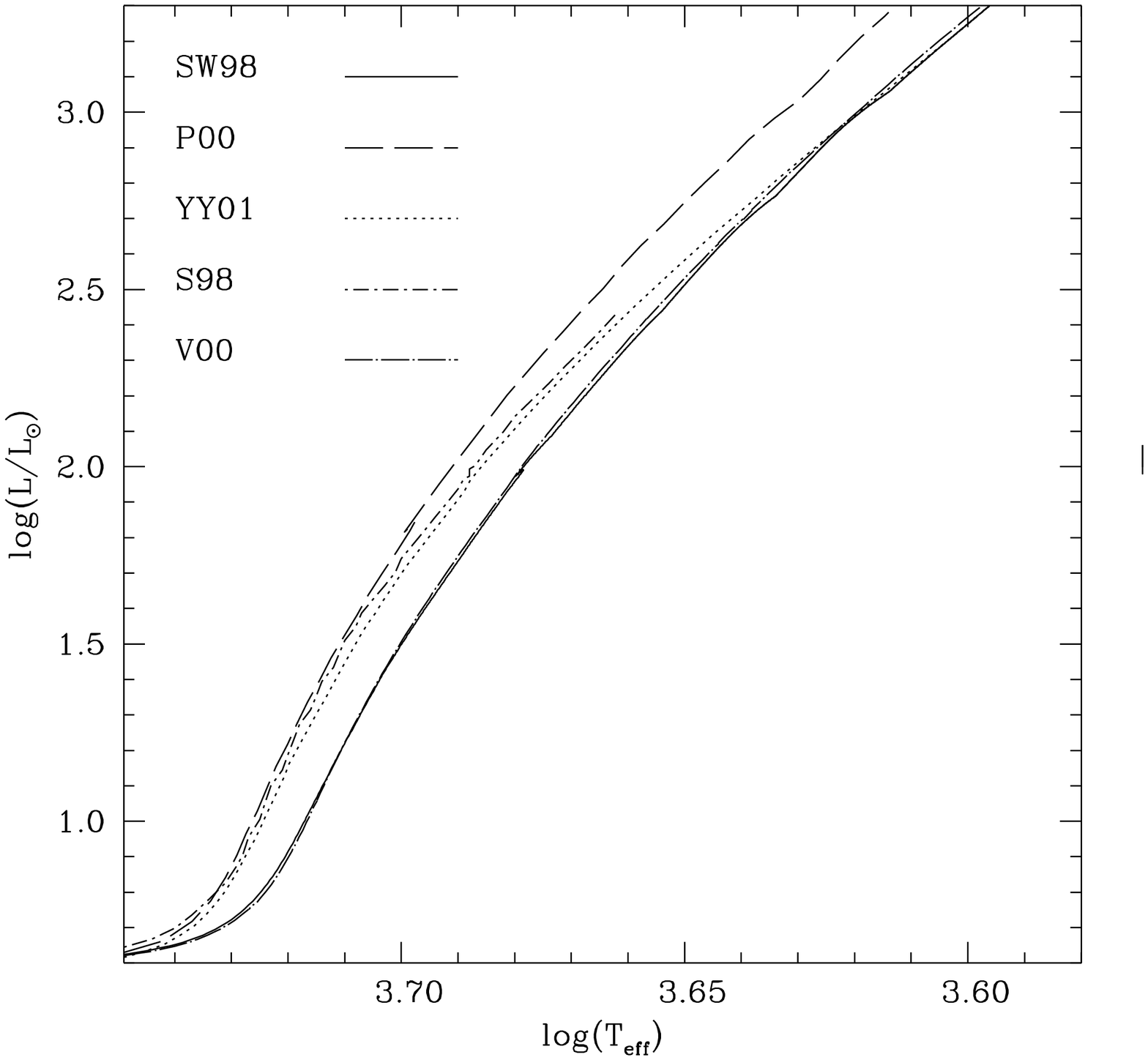}
\caption[Short caption for figure 2]{\label{fig:2} 
As Fig.~\ref{fig:1}, but provided by various authors: \cite[][POO]{Girardi00} , 
\cite[][YY01]{Yi01}, \cite[][V00]{VandenBerg00}, \cite[][SW98]{Salaris98}, 
and \cite[][S98]{Silvestri98}.}
\end{center}
\end{minipage}
\end{figure}

{\sl The outer boundary conditions:}
the procedure commonly used in the current generation of stellar models is the integration of the
atmosphere by using a functional (semi-empirical or theoretical) relation between the temperature and the
optical depth ($T(\tau)$). Recent studies of the effect of using boundary   
conditions from model atmospheres are in \cite{Montalban01} and \cite{VandenBerg08}. In Fig.~\ref{fig:1} it is shown the effects on RGB stellar models 
of different $T(\tau)$ relations, namely, the \cite{Krishna66} solar T$(\tau)$  
relationship, and the gray one. One notices that RGBs computed with a gray  T$(\tau)$ are 
systematically hotter by $\sim$100 K. In the same Fig.~\ref{fig:1}, 
we show also a RGB computed using boundary conditions from the K92 model   
atmospheres, taken at $\tau$=10.  The three displayed RGBs, for consistency,   
have been computed by employing the same low-T opacities, namely the   
ones provided by K92, in order to be homogeneous with the model atmospheres.   
The model atmosphere RGB shows a slightly different slope, crossing over the  track of the models computed with the 
\cite{Krishna66} solar T$(\tau)$, but the difference with respect to the latter stays always within   
$\sim \pm$50 K. 

Even if it is, in principle, more rigorous the use of boundary conditions provided by model atmospheres,
one has also to bear in mind that the convection treatment in the   
adopted model atmospheres \citep{Montalban01} is usually not the same as in the underlying   
stellar models (i.e., a different mixing length formalism is used).
 
{\sl Superadiabatic convection:} we already noticed that the value of $\alpha_{\rm MLT}$   
is usually calibrated by reproducing the solar $T_{\rm eff}$, and this   
solar-calibrated value is then used for stellar models of different masses and along different evolutionary phases,
including the RGB one. 
The adopted procedure guarantees that the models always predict correctly the $T_{\rm eff}$  
of at least solar type stars. However, 
the RGB location is much more sensitive to the value of $\alpha_{\rm MLT}$   
than the main sequence. This is due to the evidence that along the RGB the extension (in radius) of the 
superadiabatic layers is quite larger when compared with the MS evolutionary phase.
Therefore, it is important to verify that a solar $\alpha_{\rm MLT}$ is always suitable also for RGB stars of   
various metallicities.   
An independent way of calibrating $\alpha_{\rm MLT}$ for RGB stars is to compare empirically determined 
RGB $T_{\rm eff}$ values for GGCs with RGB models of the appropriate chemical   
composition. This kind of comparison has been performed for many of the most updated stellar models databases,
and the obtained results usually seem to suggest that the solar $\alpha_{\rm MLT}$ value is adequate also for RGB stars.    
This notwithstanding, a source of concern about an {\sl a priori} assumption of a solar $\alpha_{\rm MLT}$   
for RGB computations comes from the fact that recent models from various authors, all using a suitably     
calibrated solar value of $\alpha_{\rm MLT}$, do not show the same RGB temperatures.    
This means that -- for a fixed RGB temperature scale --   
the calibration of $\alpha_{\rm MLT}$ on the empirical $T_{\rm eff}$ values   
would not provide always the solar value \citep[see the discussion in][]{Salaris02}.   
Figure~\ref{fig:2} displays several isochrones produced by different  
groups, all computed with the same initial chemical composition, same opacities,    
and the appropriate solar calibrated values of $\alpha_{\rm MLT}$: the \cite{VandenBerg00} and \cite{Salaris98}
models are identical, the Padua ones \citep{Girardi00} are systematically hotter 
by $\sim$200 K, while the $Y^2$ ones \citep{Yi01} have a different shape.  
This comparison shows clearly that if one set of MLT solar calibrated RGBs  
can reproduce a set of empirical RGB temperatures, the others cannot,  
and therefore in some case a solar calibrated $\alpha_{\rm MLT}$  
value may not be adequate.  
The reason for these discrepancies must be due to some    
difference in the adopted input physics  
which is not compensated by the solar recalibration of $\alpha_{\rm MLT}$.   
  
\subsection{The bump of the RGB luminosity function}
\label{bump}

The RGB luminosity function (LF) of GGCs is an important tool to test the chemical stratification inside
the stellar envelopes \citep{Renzini88}.
The most interesting feature of the RGB LF is the occurrence of
a local maximum in the luminosity distribution of RGB stars, which
appears as a bump in the differential LF. This feature is caused by
the sudden increase of H-abundance left over
by the surface convection upon reaching its maximum inward extension at
the base of the RGB (\emph{first dredge up}) \citep[see][]{Thomas67}. 
When the advancing H-burning shell encounters this
discontinuity, its efficiency is affected, causing a temporary drop of the surface luminosity. After some time the
thermal  equilibrium is restored and the surface luminosity starts
to increase again. As a consequence, the stars cross the same
luminosity interval three times, and this occurrence shows up as a characteristic
peak in the differential LF of RGB stars. 

The brightness of the RGB bump is therefore related to the location of
this H-abundance discontinuity, in the sense that the deeper the chemical
discontinuity is located, the fainter is the bump luminosity. As a consequence, any physical inputs
and/or numerical assumption adopted in the computations, which affects the maximum
extension of the convective envelope strongly affects the bump brightness.
A detailed analysis of this issue can be found in \cite{CassisiSalaris97} and \cite{Cassisi97}. 
A comparison between the predicted bump luminosity and 
observations allows a direct check of how well theoretical models 
predict the extension of convective envelope and, then provide a benchmark for the evolutionary framework.

The parameter routinely adopted to compare observations with theory 
is the quantity $\Delta V_{\rm HB}^{\rm Bump}= V_{Bump}-V_{HB}$, that is, the V-magnitude  
difference between the RGB-bump and the horizontal branch (HB) at the RR Lyrae 
instability strip level. The most recent comparisons between \vhbb models and observations \citep[see Fig.~10 in][]{DiCecco10} seem to confirm a discrepancy at the level of $\sim$0.20~mag or possibly more
for GCs with total metallicity [M/H] below $\sim -$1.5, in the sense that the 
predicted RGB-bump luminosity is too high; the exact quantitative estimate of the 
discrepancy depending on the adopted metallicity scale.  
At the upper end of the GC metallicity range, the existence of a discrepancy depends 
on the adopted metallicity scale.
One drawback of using \vhbb as a diagnostic is that 
uncertainties in the determination of the observed HB level for GCs with blue HB morphologies 
and in theoretical predictions of the HB luminosity hamper any interpretation 
of discrepancies between theory and observations. 

An alternative avenue is offered by measuring the 
magnitude difference between the main sequence (MS) turn-off (TO) and the RGB-bump brightness \vtob$= V_{TO}-V_{bump}$, which bypasses the HB. This approach has been recently adopted by Cassisi et al.~(2011) by adopting a small sample
of GGCs. Although, an extension of this analysis 
to a larger, homogeneous sample of GCs is desirable; their results already provide clear evidence of a real \lq{over-luminosity}\rq\ of the predicted absolute magnitude of the RGB-bump, irrespective of problems with HB modeling and placement of the reference HB level in clusters with only blue HB stars. 

We wish also to note that the RGB LF bump provides other important
constraints for checking the accuracy of theoretical RGB models. In fact, both the shape and the location of the bump along the RGB LF can be used for investigating on the efficiency of a non-canonical
mixing at the border of the convective envelope \citep{Cassisi02} able to partially smooth the chemical
discontinuity.

\subsection{The brightness of the RGB tip}

The observational and evolutionary properties of stars at the Tip of the RGB (TRGB) play a pivotal role in current 
stellar astrophysical
research. The reasons are manifold: i) the mass size of the He core at the He flash 
fixes not only the TRGB brightness but
also the luminosity of the HB, ii) the TRGB brightness is one of the 
most important primary distance indicators.

As for the uncertainties affecting theoretical predictions about the TRGB brightness, it is clear that, being this quantity
fixed by the He core mass, any uncertainty affecting the predictions of 
$M_{core}^{He}$ immediately translates into an error on $\rm M_{bol}^{TRGB}$.
An exhaustive analysis of the physical parameters that affect the estimate of $M_{core}^{He}$ can be found 
in \cite{Salaris02}. Let us remember here that
the physical inputs that have the largest impact in the estimate of $M_{core}^{He}$ are the efficiency of 
atomic diffusion and the
conductive opacity. Unfortunately, no updates are available concerning a more realistic estimate of 
the real efficiency of diffusion in low-mass stars, apart from the Sun.
On the contrary, concerning the conductive opacity, large improvements have been obtained by
\cite[][P09]{Potekhin99}, and lately by \cite[][C07]{Cassisi07}. This new set represents a significant improvement 
(both in the accuracy and in the range of validity) with respect to previous estimates.

\begin{figure}
\begin{minipage}[t]{5.5cm}
\hspace*{-0.15cm}
\includegraphics[width=5.75cm,clip]{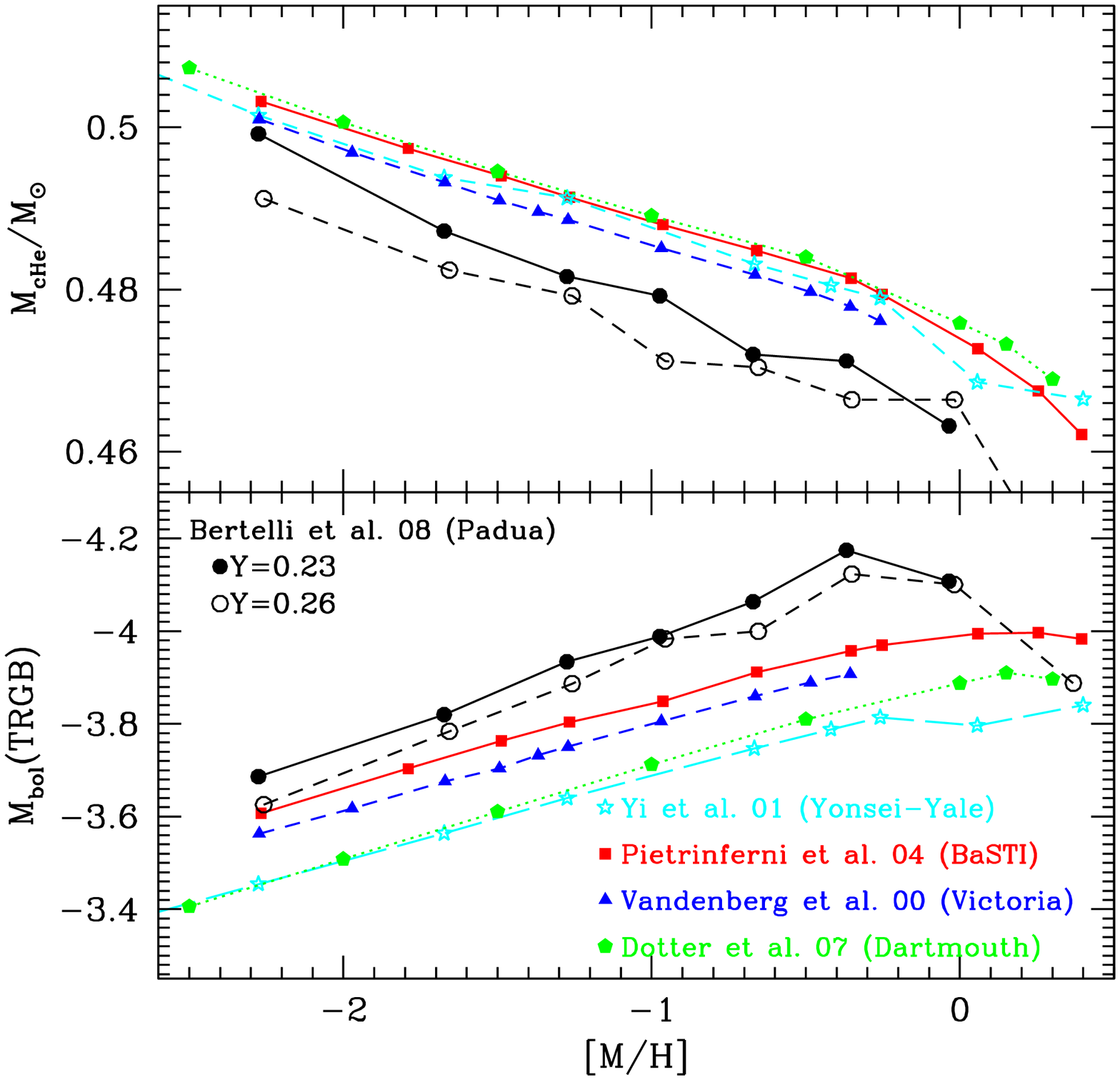}\hfill
\caption[Short caption for figure 1]{\label{fig:3} 
The trends of $M_{core}^{He}$ and $\rm M_{bol}^{TRGB}$ as a function of the metallicity as provided by 
the most recent stellar model libraries (see text for more details).}
\end{minipage}
\hfill
\begin{minipage}[t]{5.5cm}
\hspace*{-0.3cm}
\includegraphics[width=5.7cm,clip]{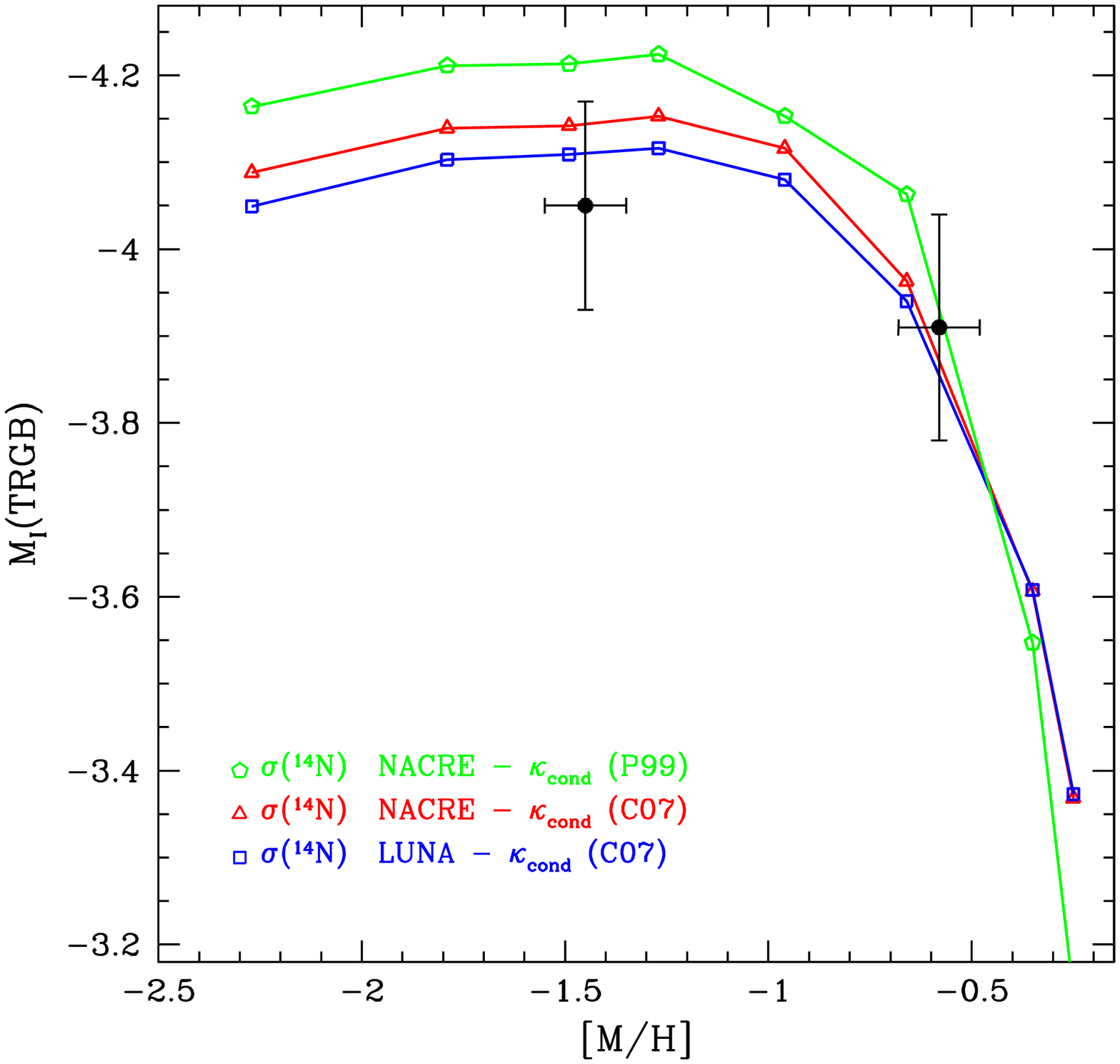}\hfill
\caption[Short caption for figure 4]{\label{fig:4} The I-band TRGB magnitude:
a comparison among GGCs data and model predictions
based on various assumptions on the conductive opacity and/or an updated $^{14}N(p,\gamma)^{15}O$ reaction rate.}
\end{minipage}
\end{figure}

We show in fig.~\ref{fig:3} the comparison of the most recent results 
(Bertelli et al. 2008 - Padua, Pietrinferni et al. 2004 - BaSTI, Vandenberg et al. 2000 - Victoria, Dotter et al. 2007 - 
Dartmouth, Yi et al. 2001 - Yonsei-Yale) concerning     
the TRGB bolometric magnitude and $M_{core}^{He}$ at the He-flash; the   
displayed quantities refer to a 0.8$M_{\odot}$ model and various   
initial metallicities.   
When excluding the Padua models, there exists a fair agreement among the various predictions about 
$M_{core}^{He}$: at fixed metallicity the 
spread among the various sets of models is at the level of $0.003M_\odot$. For the Padua models, 
we show the results corresponding to the two different initial He contents adopted by the authors: we have no clear explanation 
for the fact that the Padua models predict 
the lowest values for $M_{core}^{He}$, as well as for the presence of an \lq{erratic}\rq\ behavior of the values corresponding to the 
different He abundances: for a fixed total mass and metallicity, the $M_{core}^{He}$ value is expected to be a 
monotonic function of the initial He abundance. Concerning the trend of  $\rm M_{bol}^{\rm TRGB}$, all model 
predictions at a given metallicity are in agreement within $\sim 0.15$ mag, with the exception    
of the Padua models that appear to be brighter, at odds with the fact that they predict the lowest $M_{core}^{He}$  values.
In case of the Yonsei-Yale models, the result is also surprising since the fainter TRGB luminosity cannot be explained
by much smaller $M_{core}^{He}$ values, because this quantity is very similar to, for instance, the results given by \cite{VandenBerg00}.
When neglecting the Padua and Yonsei-Yale models, the $\sim0.1$ mag   
spread among the different TRGB brightness estimates can be interpreted in terms of differences in the adopted physical inputs.

Due to its relevance as standard candle, it is worthwhile showing a comparison between theoretical
predictions about the I-Cousins magnitude of the TRGB and empirical data. This comparison is displayed in
fig.~\ref{fig:4}, where we show the data for the GGCs  $\omega$~Cen. and 47~Tuc \citep{Bellazzini04}, and theoretical calibrations of $M_I^{TRGB}$ as a function of the metallicity based on our own
stellar models by using various assumptions concerning the conductive opacity and the rate for the nuclear reaction  $^{14}N(p,\gamma)^{15}O$ \citep[see][for details]{Pietrinferni10}. The calibrations based on the most updated physics are in fine agreement with the empirical evidence.

\section{Conclusions}

We have shown that theoretical predictions on stellar models are affected by sizeable
uncertainties, a clear proof being the occurrence of not-negligible differences between results provided by different theoretical groups. From the point of view of stellar models users, the best approach to be used for properly accounting for these uncertainties, is to not use evolutionary results with an uncritical approach and, also to adopt
as many as possible independent theoretical predictions in order to have an idea of the uncertainty existing in the match between
theory and observations. 

On the other hand, stellar model makers should continue their effort of continuously updating their models in order to account for the \lq{best}\rq\ physics available at any time, and consider the various empirical constraints as a benchmark of their stellar models. This represents a fundamental step for obtaining as much as possible accurate and reliable stellar models.

\begin{acknowledgement}    
 We warmly thank the LOC and the SOC for
organizing this interesting meeting. 
\end{acknowledgement}

\end{document}